\documentclass[pdftex,twocolumn,epjc3]{svjour3}

\usepackage{graphicx}
\usepackage{mathptmx,amsmath,amsfonts,amssymb,bigints}
\usepackage[T1]{fontenc}
\usepackage{flushend}
\usepackage[numbers,sort&compress]{natbib}
\usepackage[colorlinks,citecolor=blue,urlcolor=blue,linkcolor=blue]{hyperref}
\usepackage{subfigure}

\newcommand{\be}{\begin{equation}}
\newcommand{\ee}{\end{equation}}
\newcommand{\ben}{\begin{eqnarray}}
\newcommand{\een}{\end{eqnarray}}
\newcommand{\bei}{\begin{IEEEeqnarray}}
\newcommand{\eei}{\end{IEEEeqnarray}}
\newcommand{\bes}{\begin{subequations}}
\newcommand{\ees}{\end{subequations}}
\newcommand{\bb}{\bibitem}

\newcommand{\nn}{\nonumber\\}
\newcommand{\bfi}{\begin{figure}}
\newcommand{\efi}{\end{figure}}
\newcommand{\bc}{\begin{center}}
\newcommand{\ec}{\end{center}}

\journalname{Eur. Phys. J. C}

\begin{document}
\title{Twinlike models for parametrized dark energy}

\author{J.D. Dantas\thanksref{e1,addr1}
	\and
	   J.J. Rodrigues\thanksref{e2,addr2} }

\thankstext{e1}{e-mail: jddantas@ufcg.edu.br}

\thankstext{e2}{e-mail: jjrodrigues@uepb.edu.br}

\institute{Unidade Acad\^emica de F\'\i sica e Matem\'atica, Universidade Federal de Campina Grande, 58175-000 Cuit\'e, PB, Brazil\label{addr1}
	\and
	Departamento de F\'{\i}sica, Universidade Estadual da Para\'{\i}ba, 58233-000 Araruna, PB, Brazil,\label{addr2} }

\date{Received: date / Accepted: date}

\maketitle

\abstract{
We study cosmological models involving a single real scalar field that has an equation of state parameter which evolves with cosmic time. We highlight some common parametrizations for the equation of state as a function of redshift in the context of twinlike theories. The procedure is used to introduce different models that have the same acceleration parameter, with the very same energy densities and pressure in flat spacetime.}



\section{Introduction}

Cosmological observations of Type Ia Supernovae and Cosmic Microwave Background suggest that the Universe had started to accelerate its expansion at the present epoch \cite{perlmutter1,riess1,perlmutter2,riess2,wmap9,planck18}. The standard explanation refers to an exotic component which has positive energy density and negative pressure, known as ``dark energy'' (DE). A variety of theoretical models has been proposed to explain this acceleration. The most natural and simplest model for DE is the $\Lambda$CDM model, containing a mixture of cosmological constant $\Lambda$ and cold dark matter (CDM), for which the equation of state parameter is $\omega=-1$ \cite{peebles1,padma,carrol}. However, this model suffers from two major problems, namely, fine-tuning and cosmological coincidence problems \cite{carrol,weinberg,steinhardt}. 

In order to solve these problems, alternative DE models have been proposed, where the equation of state parameter evolves with cosmic time, mainly the canonical scalar field DE models \cite{peebles1,efstathiou,weller,bassett} - see also \cite{peebles2,peebles3} (for a historical review), which study the cosmological evolution with minimally coupled scalar fields. Another possibility is to consider non-canonical scalar field models, which has shown increasing interest \cite{planckxiv,padma1,jassal,linder1,mamon1,mamon2}. We pay special attention on the tachyonic fields, which emerged in the context of string theories \cite{sen1,sen2,sen3,garousi}, and have been intensively studied in cosmology \cite{padma1,gibbons,frolov,bagla,abramo}. It is possible to have accelerated expansion of the universe during the late times for both choices, and we have searched for situations in which correspondences can be established from a modified potential function.

The correspondence above referred has already been investigated for defect structures, describing different scalar field theories with very similar properties \cite{andrews,bazeia1,adam1,bazeia2,adam2,bazeia3,bazeia4,bazeia5}. The same idea was applied in the context of Friedmann-Robertson-Walker (FRW) cosmology \cite{bazeia6}, in a previous paper. In the present work, we extend this view and consider some popular DE parametrizations for canonical and tachyonic scalar field models. We find that the two models present the very same acceleration parameter, with the same energy density, and we name them twinlike models. 

The canonical potential and the tachyonic potential are distinct, but they lead to the same cosmological evolution. Our motivation for studying these twinlike models is as follows: the medium with negative pressure capable of accelerating the expansion of the universe has two possible sources - a minimally coupled scalar field (canonical) or a non-minimally coupled scalar field (tachyonic).

The basic concepts are presented in Sec. \ref{eeq}. In Sec. \ref{ntwin} we investigated the twin nature of the standard and tachyonic models. In Sec. \ref{ilust} we present some illustrations. The paper ends with a summary in Sec. \ref{sandc}.


\section{Einstein equations}
\label{eeq}

In order to investigate this proposal, we present some basic theoretical considerations. The action for a universe with spacetime curvature $R$, filled with a scalar field $\phi$ and containing matter, is given by
\be
S=\bigintssss d^4x\sqrt{-g}\left[-\frac{1}{4}R+{\cal L}(\phi,X)\right]+S_m
\ee
where we have made $4\pi G=c=1$, $X=\frac{1}{2}\partial^{\mu}\phi\partial_{\mu}\phi$, and $S_m$ is the action of the matter.

The metric representing a homogeneous, isotropic and spatially flat universe is the FRW metric
\be
ds^2=dt^2-a^2(t)\left(dr^2+r^2d\theta^2+r^2\sin^2\theta\;d\phi^2\right)
\ee
where $a(t)$ is the scale factor of the universe, $r$ is the radial coordinate and $d\Omega^2=d\theta^2+\sin^2\theta\;d\phi^2$ describes the angular portion of the metric. In this scenario, the Einstein equations are
\be
H^2=\frac{2}{3}(\rho_{\phi}+\rho_m)
\label{h2}
\ee
\be
\dot{H}=-(\rho_{\phi}+p_{\phi}+\rho_m)
\ee
where $\rho_{\phi}$ and $p_{\phi}$ are respectively energy density and pressure of the scalar field $\phi$, $\rho_m$ represents the energy density of the matter component of the universe, $H=\dot{a}/a$ denotes Hubble parameter, and an overdot indicates differentiation with respect to time $t$.

The conservation of the scalar field and matter is represented respectively by the equations of continuity
\be
\dot{\rho}_{\phi}+3H(\rho_{\phi}+p_{\phi})=0
\label{conti}
\ee
\be
\dot{\rho}_m+3H\rho_m=0
\label{contim}
\ee
and the cosmic acceleration parameter is given by
\be
q=\frac{\ddot{a}a}{\dot{a}^2}=1+\frac{\dot{H}}{H^2}
\ee

Rewriting the equations in terms of redshift $z=\dfrac{a_0}{a}-1$, from (\ref{conti}) and (\ref{contim}), we obtain
\be
\rho_{\phi}(z)=\rho_{\phi 0}\exp\left(3\int_0^z\dfrac{1+\omega_{\phi}(z')}{1+z'}dz'\right)
\label{densidade}
\ee
\be
\rho_m(z)=\rho_{m0}(1+z)^3
\label{densidadem}
\ee
where $\omega_{\phi}=p_{\phi}/\rho_{\phi}$ is the dark energy EoS parameter and the subscript $0$ indicates the present epoch. The Friedmann equations then take the form

\ben
H^2&=&H_0^2\bigg[\Omega_{m0}(1+z)^3+\nn
&+&\Omega_{\phi 0}\exp\left(3\int_0^z\dfrac{1+\omega_{\phi}(z')}{1+z'}dz'\right)\bigg]
\een

\ben
\dot{H}&=&-\dfrac{3}{2}H_0^2\bigg[\Omega_{m0}(1+z)^3+\nn
&+&\Omega_{\phi0}\left(1+\omega_{\phi}(z)\right)\exp\left(3\int_0^z\dfrac{1+\omega_{\phi}(z')}{1+z'}dz'\right)\bigg]
\een
where $\Omega_{m0}=\dfrac{2\rho_{m0}}{3H_0^2}$ and $\Omega_{\phi0}=\dfrac{2\rho_{\phi0}}{3H_0^2}=1-\Omega_{m0}$ are the density parameters of the matter and scalar field, respectively, at the present epoch. The acceleration parameter is also rewritten as
\be
q=1-(1+z)\dfrac{d\ln H(z)}{dz}
\label{parametro}
\ee


\section{The twinlike models}
\label{ntwin}


\subsection{Standard case}

If the scalar field (dark energy) is described by the standard dynamics, we have
\be
{\cal L}=X-V(\phi)
\ee
where $V(\phi)$ is the potential of the scalar field. In this case, energy density and pressure are given by
\be
\rho_{\phi}=\frac{1}{2}\dot{\phi}^2+V(\phi)
\label{repstandard1}
\ee
\be
p_{\phi}=\frac{1}{2}\dot{\phi}^2-V(\phi)
\label{repstandard2}
\ee
and the scalar field evolves as follows
\be
\ddot{\phi}+3H\dot{\phi}+V_{,\phi}=0
\ee

From (\ref{repstandard1}) and (\ref{repstandard2}), we express the potential,
\be
V(z)=\dfrac{1}{2}\left[1-\omega_{\phi}(z)\right]\rho_{\phi}(z)
\ee
and we write an equation for the scalar field,
\be
\phi_{,z}=\dfrac{\sqrt{\left[1+\omega_{\phi}(z)\right]\rho_{\phi}(z)}}{(1+z)H}
\label{dpstandard}
\ee
both in terms of redshift $z$.


\subsection{Tachyonic modified case}
Let us now consider the scalar field described by tachyonic dynamics. We change the model as
follows
\be
{\cal L}=-U(\phi)\sqrt{1-2X}+f(\phi)
\ee
where $U(\phi)$ and $f(\phi)$ are functions to be determined. Energy density and pressure are now given by
\be
\rho_{\phi}=\frac{U}{\sqrt{1-\dot{\phi}^2}}-f
\label{repmodified1}
\ee
\be
p_{\phi}=-U\sqrt{1-\dot{\phi}^2}+f
\label{repmodified2}
\ee
and the scalar field obeys
\be
\ddot{\phi}+\left(1-\dot{\phi}^2\right)\left(3H\dot{\phi}+\frac{U_{,\phi}}{U}\right)-\left(1-\dot{\phi}^2\right)^{3/2}\frac{f_{,\phi}}{U}=0
\ee

From (\ref{repmodified1}) and (\ref{repmodified2}), in terms of redshift $z$, we obtain
\be
U(z)=\sqrt{\left[\rho_{\phi}(z)+f\right]\left[f-\rho_{\phi}(z)\omega_{\phi}(z)\right]}
\ee
and
\be
\phi_{,z}=\dfrac{1}{(1+z)H}\sqrt{\dfrac{\left[1+\omega_{\phi}(z)\right]\rho_{\phi}(z)}{f+\rho_{\phi}(z)}}
\label{dpmodified}
\ee


\subsection{The twin nature}

In order to get to twinlike models, we need to make the appropriate choice for $f(z)$. In this case we consider
\be
f(z)=1-\rho_{\phi}(z)
\ee
So, the modified potential takes the form
\be
U=\sqrt{1-\left[1+\omega_{\phi}(z)\right]\rho_{\phi}(z)}
\ee

In both cases (standard and modified), we have the same energy density, given by (\ref{densidade}), and the same pressure. The scalar field is also the same in both cases, being
\be
\phi(z)=\phi_0+\bigintss_0^z{\dfrac{\sqrt{\left[1+\omega_{\phi}(z')\right]\rho_{\phi}(z')}}{(1+z')H(z')}}dz'
\label{phidz}
\ee
The acceleration parameter also has the same form, given by (\ref{parametro}), and the Friedmann equations have the same evolution in both cases. The models are twin. 

However, twin models can be further differentiated! The fingerprint signature is defined by the effective speed of sound, entering a general rule for the evolution of small perturbations \cite{erickson}. We can obtain:
\be
c_s^2=\frac{p_{\phi,X}}{\rho_{\phi,X}}=\frac{{\cal L}_{,X}}{{\cal L}_{,X}+2{\cal L}_{,XX}X}
\ee
Disregarding the trivial solution $f(\phi)=U(\phi)\sqrt{1-2X}$, when ${\cal L}=0$, the speed of sound can evolve differently for the chosen model, and we can admit the solution $c_s^2<1$, which leads to the growth of inhomogeneities in the present cosmic acceleration - see \cite{bean,blomqvist} for a further discussion. In this sense, the growth of inhomogeneities can occur differently for twin models, but with no change in evolution of the density, or acceleration parameter, of the Universe as a whole. Explicitly, twin models describe the same cosmic expansion, being able to measure local changes in the growth of inhomogeneities.

In the next section we present how to build twinlike models for dark energy models. It is important to emphasize that the results presented are valid for the current acceleration regime. In the context of the primordial universe, it is also possible to obtain twinlike models, see \cite{adam3}, where the slow-roll inflation, evolving under different potentials, lead to a very similar inflationary phase.   

\section{Illustrations}
\label{ilust}


\subsection{Cosmological constant}

As a first example we take $\omega_{\phi}(z)=\omega_0$, a cosmological constant, in the limit $-1<\omega_0<-\dfrac{1}{3}$. In this situation, the energy density of scalar field is written as
\be
\rho_{\phi}(z)=\dfrac{3}{2}H_0^2(1-\Omega_{m0})\left(1+z\right)^{3\left(1+\omega_0\right)}
\label{rhodz1}
\ee
So the potentials of standard and modified cases are, respectively,
\be
V(z)=\dfrac{3}{4}H_0^2(1-\Omega_{m0})\left(1-\omega_0\right)\left(1+z\right)^{3\left(1+\omega_0\right)}
\label{v1}
\ee
\be
U(z)=\dfrac{1}{2}\sqrt{4-6H_0^2(1-\Omega_{m0})\left(1+\omega_0\right)\left(1+z\right)^{3\left(1+\omega_0\right)}}
\label{u1}
\ee
The evolution of the Hubble parameter with the redshift is given by the Friedmann equation
\be
H^2(z)=H_0^2\left[\Omega_{m0}(1+z)^3+(1-\Omega_{m0})(1+z)^{3(1+\omega_0)}\right]
\ee
And the acceleration parameter is
\be
q(z)=-\dfrac{1}{2}-\dfrac{3}{2}\left[\dfrac{\omega_0}{1+\alpha(1+z)^{-3\omega_0}}\right]
\ee
where $\alpha=\Omega_{m0}/(1-\Omega_{m0})$.

The density parameters of the matter and scalar field are respectively
\be
\Omega_m(z)=\dfrac{1}{1+\frac{1}{\alpha}(1+z)^{3\omega_0}}
\ee

\be
\Omega_{\phi}(z)=\dfrac{1}{1+\alpha(1+z)^{-3\omega_0}}
\ee
With the help of Eq. (\ref{rhodz1}), we can solve Eq. (\ref{phidz}) for the scalar field analytically. The result is
\ben
\phi(z)&=&\phi_0+\dfrac{\sqrt{6(1+\omega_0)}}{3\omega_0}\bigg[\tanh^{-1}\left(\sqrt{1+\alpha(1+z)^{-3\omega_0}}\right)-\nn
&-&\tanh^{-1}\left(\sqrt{1+\alpha}\right)\bigg]
\een

\begin{figure*}[htb!]
\centering
\subfigure[]{\includegraphics[scale=0.39]{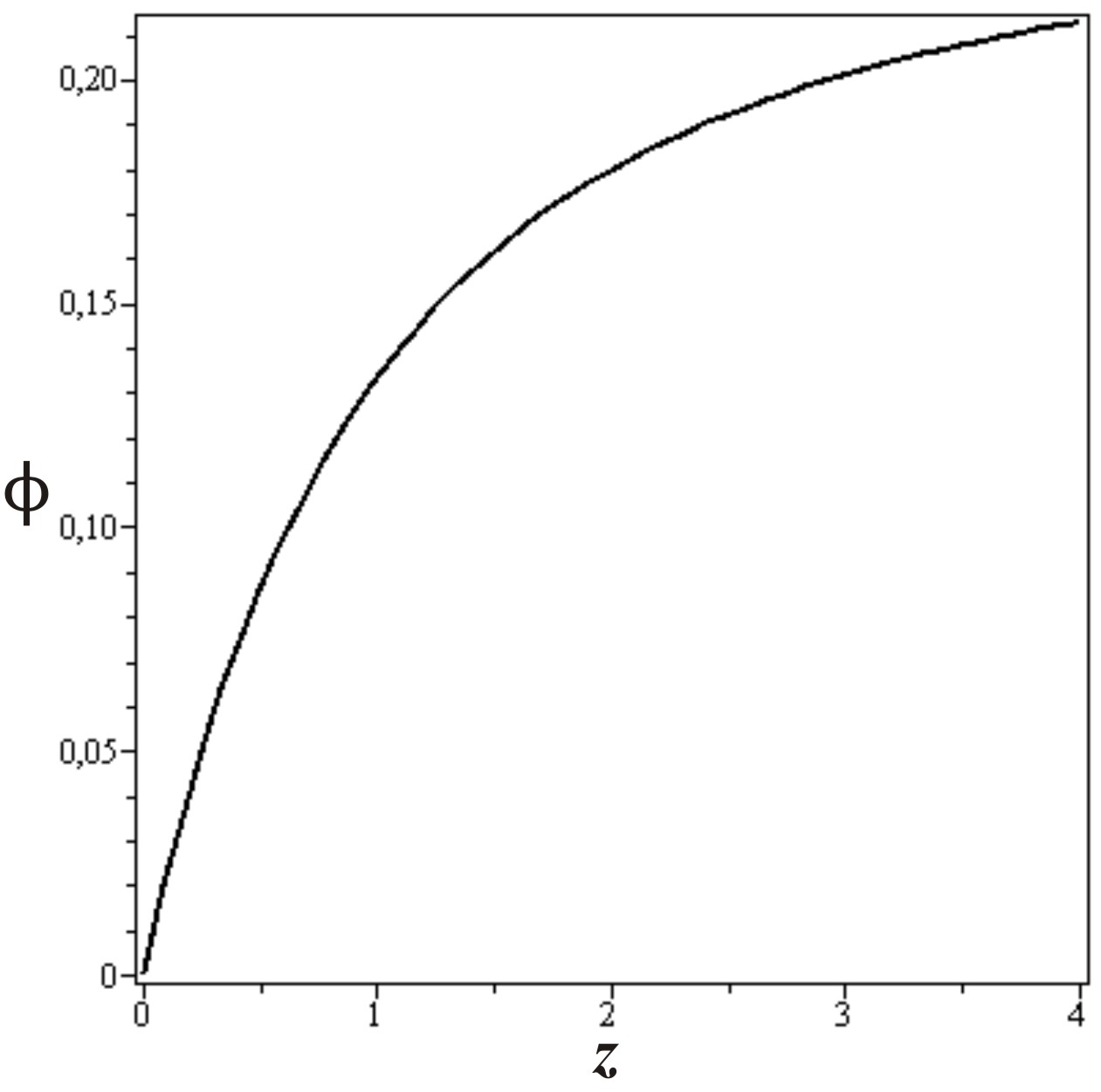}}\;\subfigure[]{\includegraphics[scale=0.39]{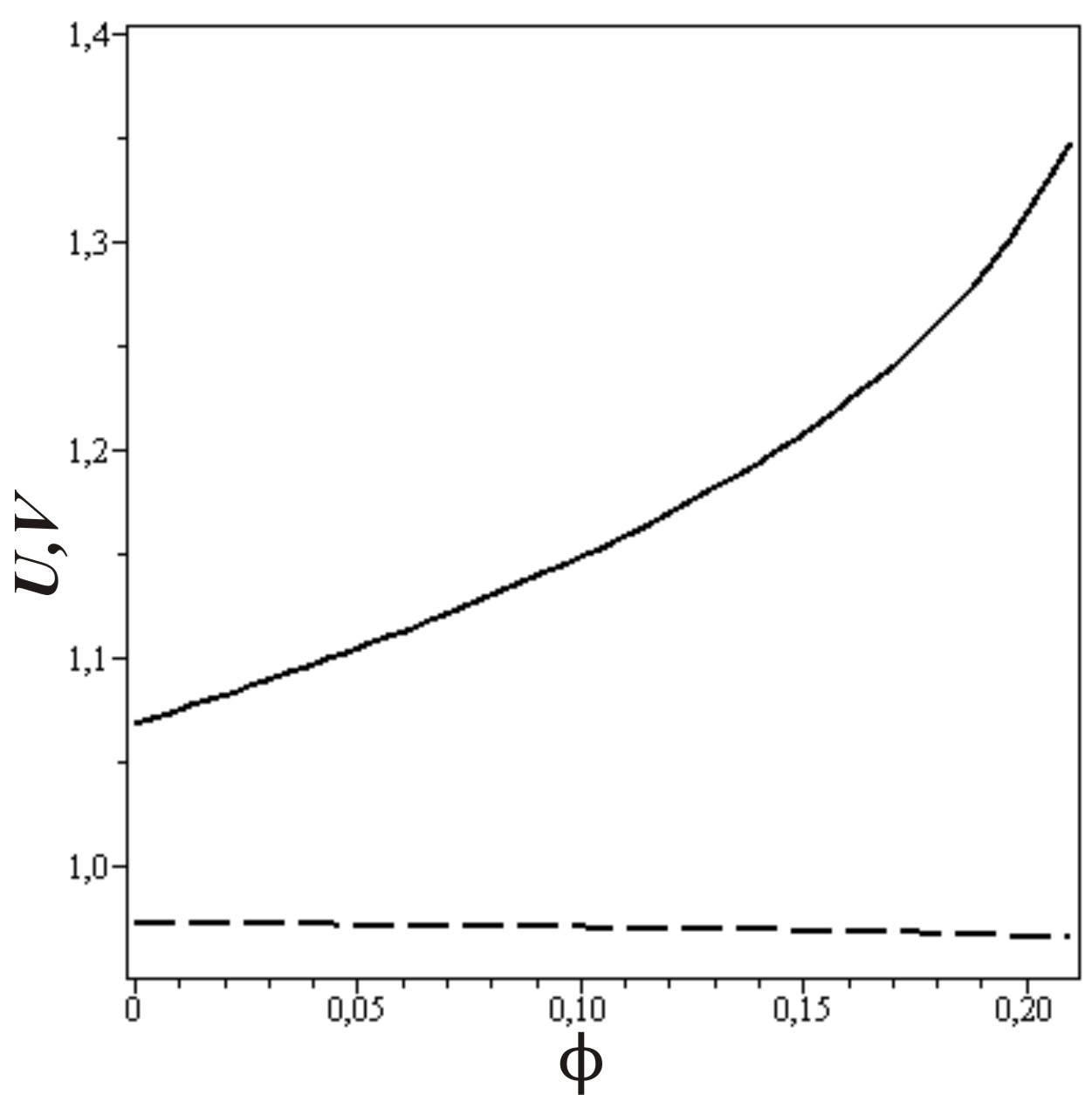}}\;\subfigure[]{\includegraphics[scale=0.39]{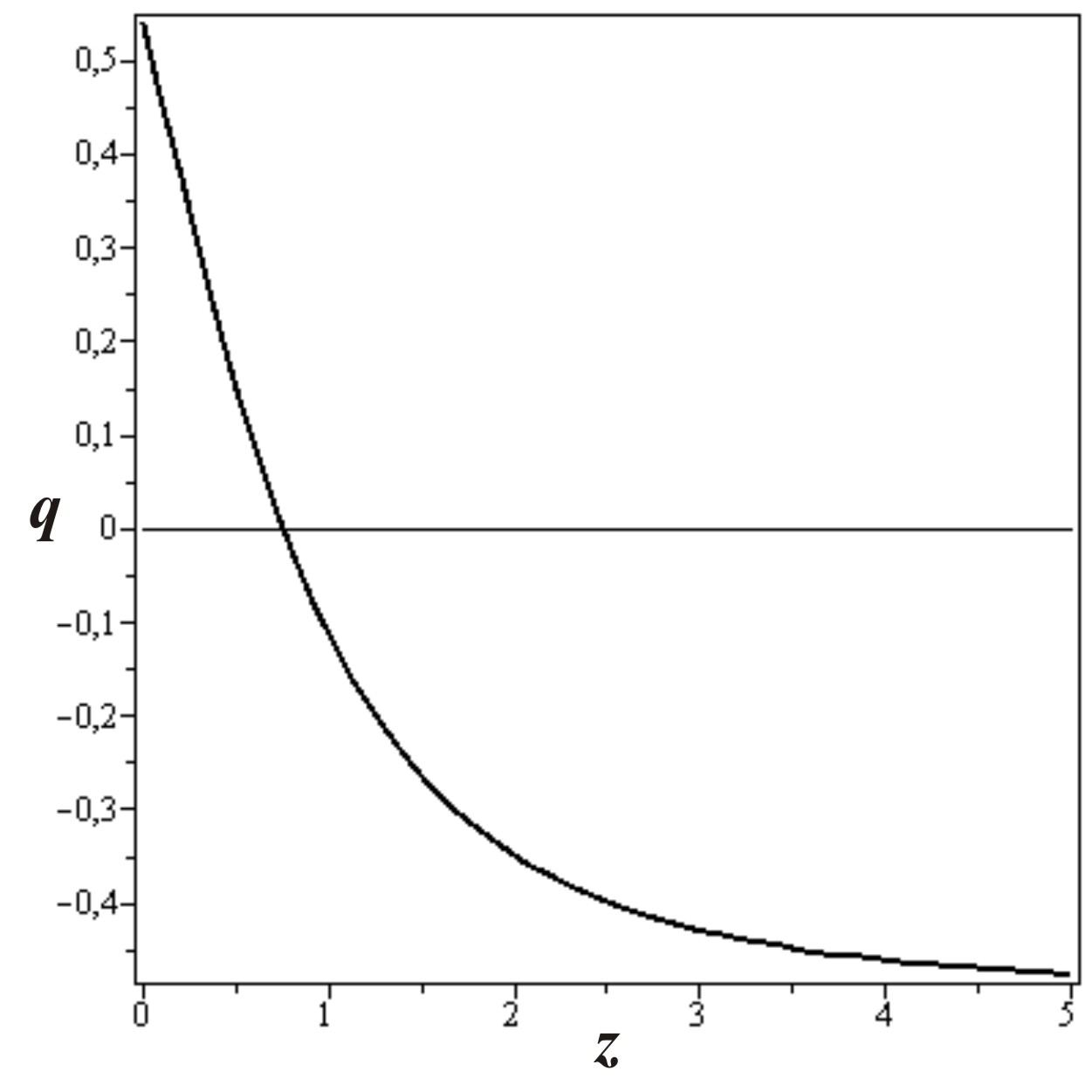}}\;\subfigure[]{\includegraphics[scale=0.39]{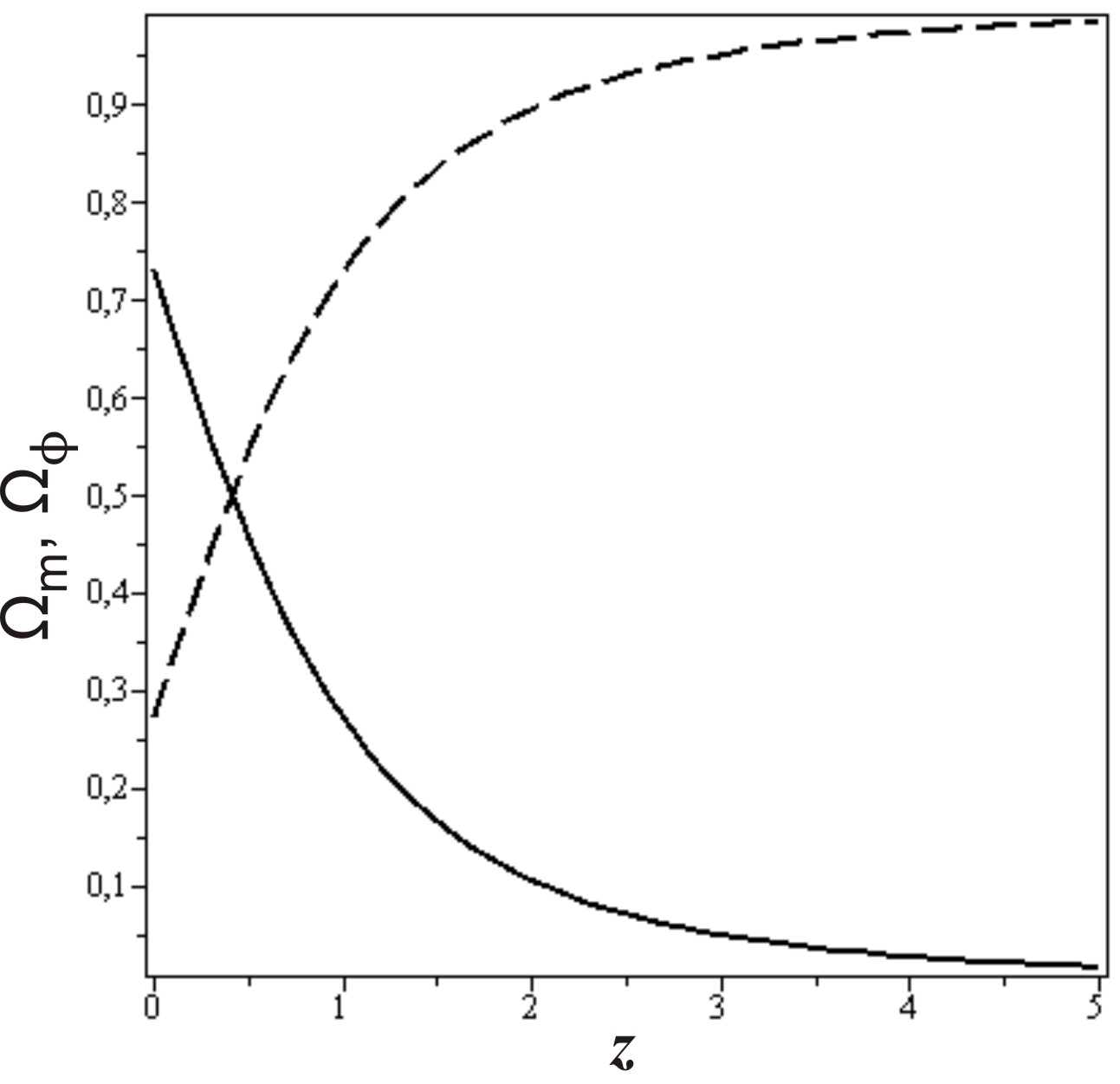}}
\caption{(a) Plot of $\phi$ as a function of $z$. (b) Plot of $V$ (solid curve) and $U$ (dashed curve) as a function of $\phi$. (c) Plot of $q$ as a function of $z$. (d) Plot of $\Omega_{\phi}$ (solid curve) and $\Omega_m$ (dashed curve) as a function of $z$. This is for cosmological constant, with $\phi_0=0$, $H_0=1$, $\omega_0=-0.95$ and $\Omega_{m0}=0.27$.}
\label{fconstant}
\end{figure*}

Figure \ref{fconstant}(a) shows the plot of $\phi(z)$. Equations (\ref{v1}) and (\ref{u1}) express $V$ and $U$ as functions of $z$. It is very difficult to work with these potentials in terms of $\phi$. Figure \ref{fconstant}(b) shows the plot of $V(\phi)$ and $U(\phi)$ from numerical results. The $V$ and $U$ curves are clearly distinct, but the twin nature is shown in the graph of the acceleration parameter $q$, which is the same for both models. The plot of $q(z)$ in Figure \ref{fconstant}(c) shows the transition from a decelerating to an accelerating regime as $z$ decreases. The evolutions of $\Omega_{\phi}$ and $\Omega_m$ are showns in Figure \ref{fconstant}(d). Note that $\Omega_{\phi}$ starts dominating over $\Omega_m$ at around $z\sim0.4$.

\subsection{Linear parametrization}

As a second example we now consider $\omega_{\phi}(z)=\omega_0+\omega_1z$ \cite{cooray1,weller1}. In this case, the energy density of scalar field takes the form
\be
\rho_{\phi}(z)=\dfrac{3}{2}H_0^2(1-\Omega_{m0})\left(1+z\right)^{3\left(1+\omega_0-\omega_1\right)}\exp(3\omega_1z)
\ee
So the potentials of standard and modified cases are, respectively,
\ben
V(z)&=&\dfrac{3}{4}H_0^2(1-\Omega_{m0})\big(1-\omega_0-\nn
&-&\omega_1z\big)\left(1+z\right)^{3(1+\omega_0-\omega_1)}\exp\left(3\omega_1z\right) \label{v2}
\een

\ben
U(z)&=&\dfrac{1}{2}\Big[4-6H_0^2(1-\Omega_{m0})\big(1+\omega_0+\nn
&+&\omega_1z\big)\left(1+z\right)^{3(1+\omega_0-\omega_1)}\exp\left(3\omega_1z\right)\Big]^{1/2}
\label{u2}
\een
The Friedmann equation is given by
\ben
H^2(z)&=&H_0^2\Big[\Omega_{m0}(1+z)^3+\nn
&+&(1-\Omega_{m0})(1+z)^{3(1+\omega_0-\omega_1)}\exp\left(3\omega_1z\right)\Big]
\een
The acceleration parameter is
\be
q(z)=-\dfrac{1}{2}-\dfrac{3}{2}\left[\dfrac{\omega_0+\omega_1z}{1+\alpha(1+z)^{-3(\omega_0-\omega_1)}\exp(-3\omega_1z)}\right]
\ee
The density parameters of the matter and scalar field are, respectively,
\be
\Omega_m(z)=\dfrac{1}{1+\frac{1}{\alpha}(1+z)^{3(\omega_0-\omega_1)}\exp(3\omega_1z)}
\ee

\be
\Omega_{\phi}(z)=\dfrac{1}{1+\alpha(1+z)^{-3(\omega_0-\omega_1)}\exp(-3\omega_1z)}
\ee

\begin{figure*}[htb!]
\centering
\subfigure[]{\includegraphics[scale=0.39]{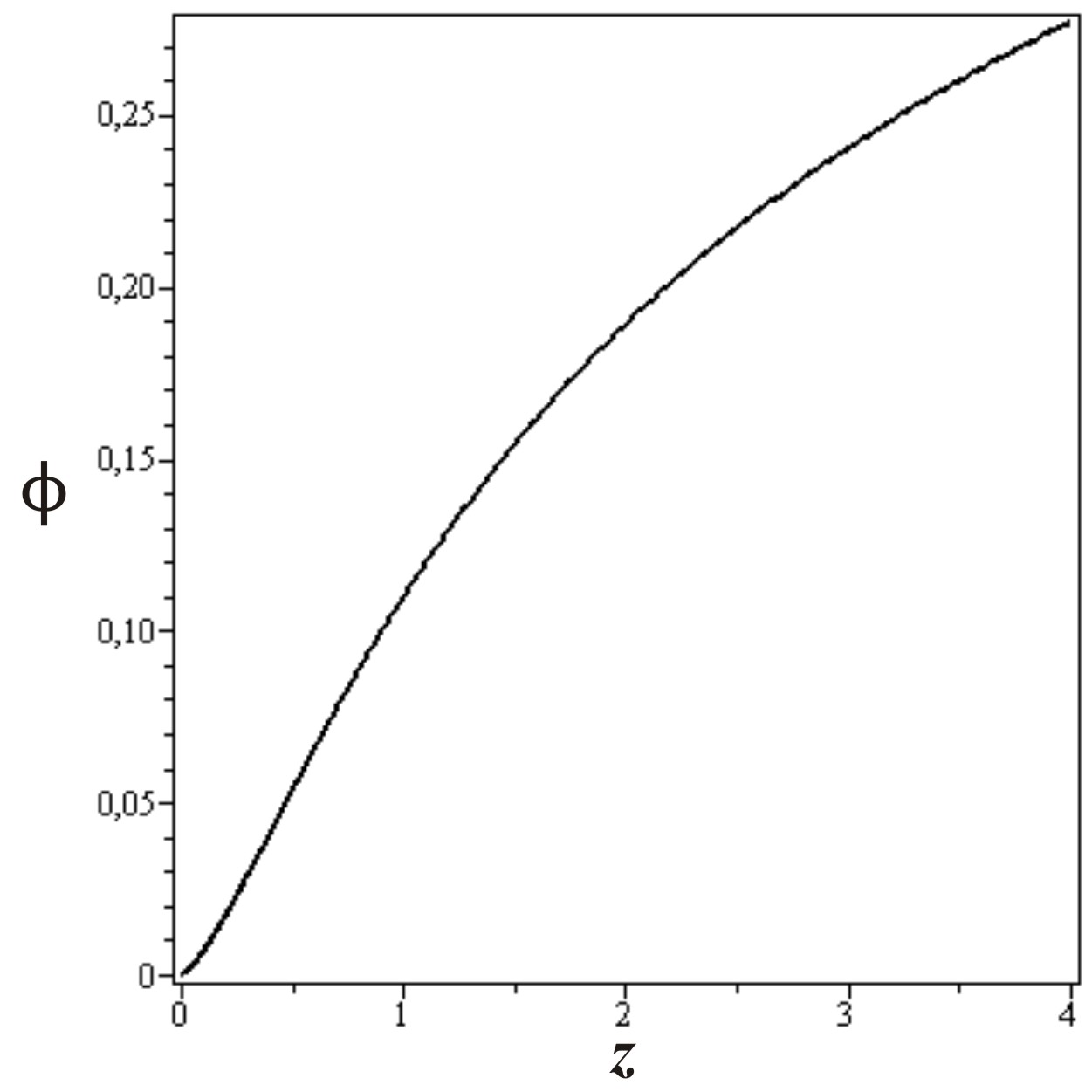}}\;\subfigure[]{\includegraphics[scale=0.39]{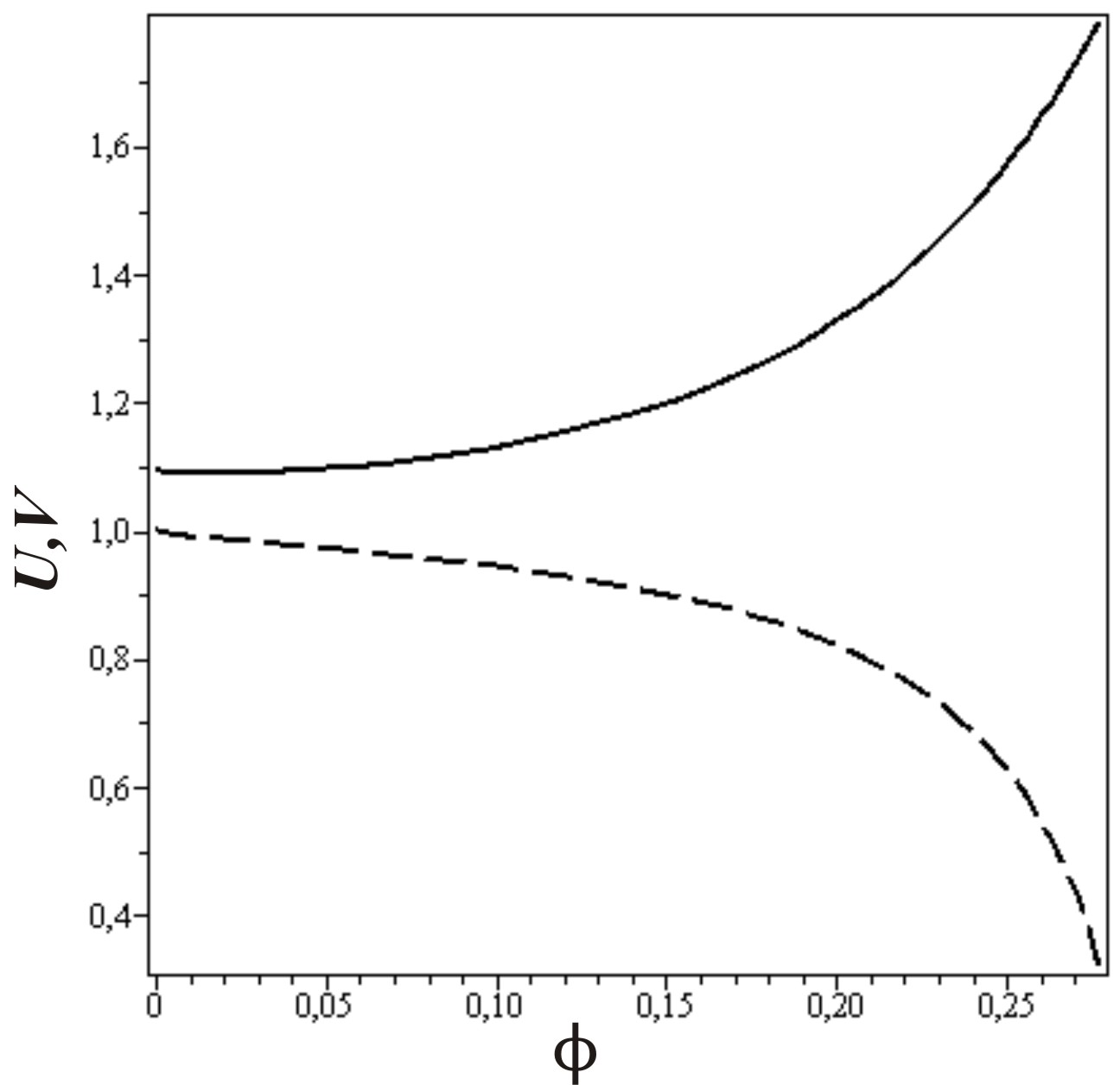}}\;\subfigure[]{\includegraphics[scale=0.39]{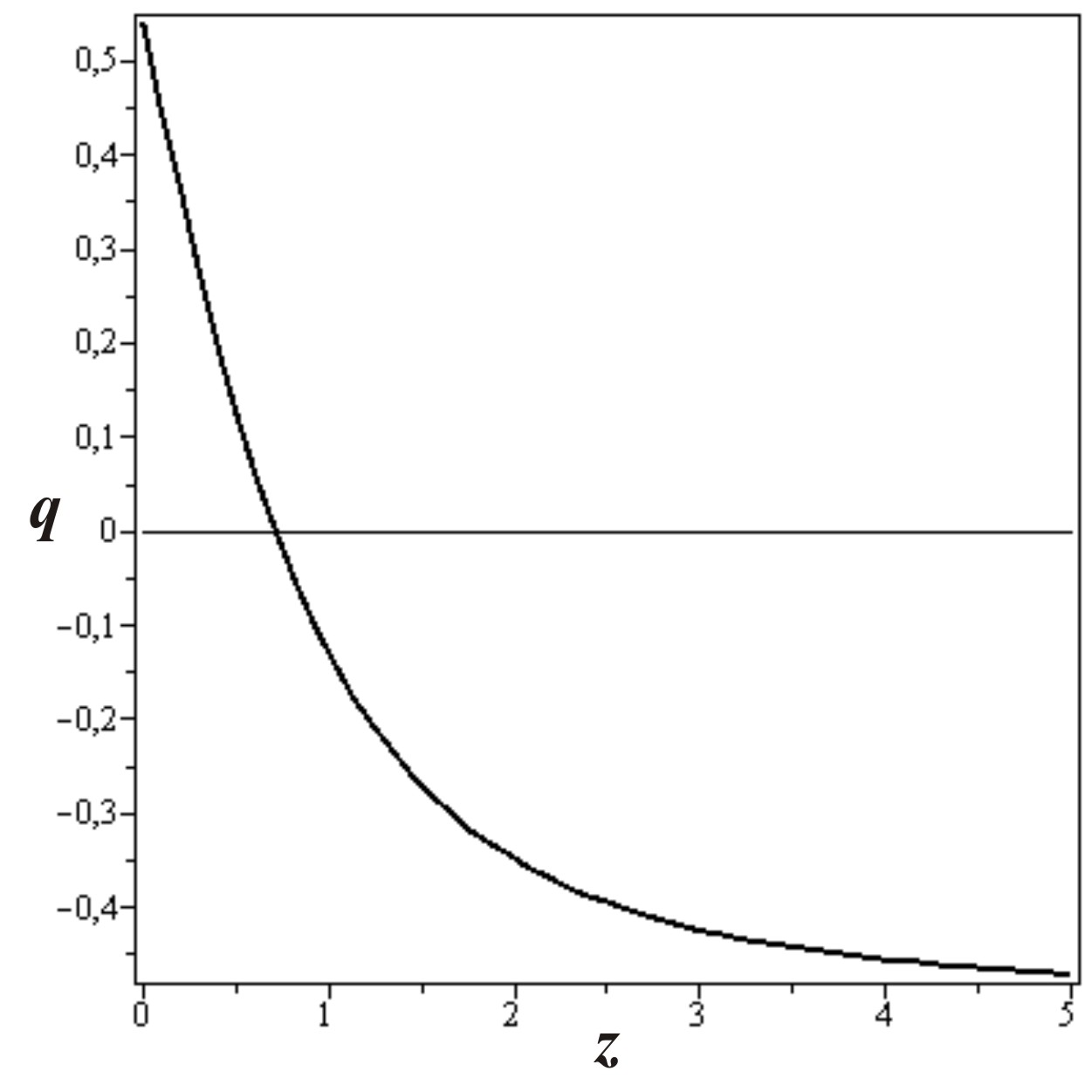}}\;\subfigure[]{\includegraphics[scale=0.39]{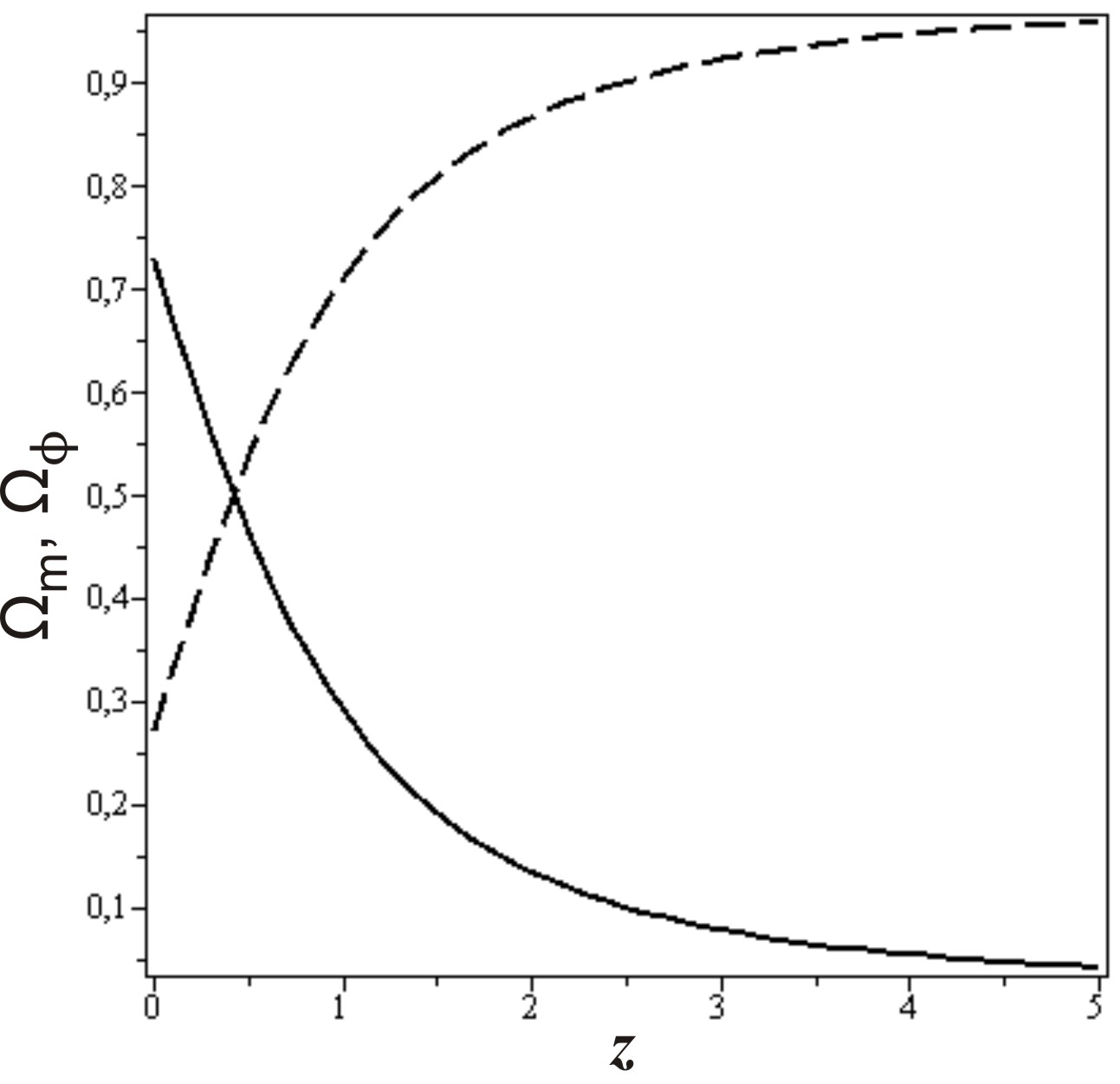}}
\caption{(a) Plot of $\phi$ as a function of $z$. (b) Plot of $V$ (solid curve) and $U$ (dashed curve) as a function of $\phi$. (c) Plot of $q$ as a function of $z$. (d) Plot of $\Omega_{\phi}$ (solid curve) and $\Omega_m$ (dashed curve) as a function of $z$. This is for linear parametrization, with $\phi_0=0$, $H_0=1$, $\omega_0=-1$, $\omega_1=0.1$ and $\Omega_{m0}=0.27$.}
\label{flinear}
\end{figure*}

Figure \ref{flinear} shows the plots of (a) $\phi(z)$, (b) $V(\phi)$ and $U(\phi)$ from numerical results. The plot of $q(z)$ in Figure \ref{flinear}(c) shows also the transition from a decelerating to an accelerating regime as $z$ decreases. The evolutions of $\Omega_{\phi}$ and $\Omega_m$ are shown in Figure \ref{flinear}(d), and $\Omega_{\phi}$ starts dominating over $\Omega_m$ at around $z\sim0.4$.

\subsection{Chevallier-Polarski-Linder (CPL) parametrization}

The CPL parametrization \cite{chevallier, linder2, jing, scherrer} is characterized by
\be
\omega_{\phi}(z)=\omega_0+\omega_1\left(\dfrac{z}{1+z}\right)
\ee
The energy density of scalar field is

\be
\rho_{\phi}(z)=\dfrac{3}{2}H_0^2(1-\Omega_{m0})\left(1+z\right)^{3\left(1+\omega_0+\omega_1\right)}\exp\left(-\frac{3\omega_1z}{1+z}\right)
\ee
The potentials of standard and modified cases are, respectively,
\ben
V(z)&=&\dfrac{3}{4}H_0^2(1-\Omega_{m0})\bigg(1-\omega_0-\nn
&-&\frac{\omega_1z}{1+z}\bigg)\left(1+z\right)^{3(1+\omega_0+\omega_1)}\exp\left(-\frac{3\omega_1z}{1+z}\right)
\label{v2}
\een

\ben
U(z)&=&\dfrac{1}{2}\bigg[4-6H_0^2(1-\Omega_{m0})\bigg(1+\omega_0+\nn
&+&\frac{\omega_1z}{1+z}\bigg)\left(1+z\right)^{3(1+\omega_0+\omega_1)}\exp\left(-\frac{3\omega_1z}{1+z}\right)\bigg]^{1/2}
\label{u2}
\een
The Friedmann equation is
\ben
H^2(z)&=&H_0^2\bigg[\Omega_{m0}(1+z)^3+\nn
&+&(1-\Omega_{m0})(1+z)^{3(1+\omega_0+\omega_1)}\exp\left(-\frac{3\omega_1z}{1+z}\right)\bigg]
\een
The acceleration parameter is given by
\be
q(z)=-\dfrac{1}{2}-\dfrac{3}{2}\left[\dfrac{\omega_0+\omega_1\left(\frac{z}{1+z}\right)}{1+\alpha(1+z)^{-3(\omega_0+\omega_1)}\exp\left(\frac{3\omega_1z}{1+z}\right)}\right]
\ee
The density parameters of the matter and scalar field are, respectively,
\be
\Omega_m(z)=\dfrac{1}{1+\frac{1}{\alpha}(1+z)^{3(\omega_0+\omega_1)}\exp\left(-\frac{3\omega_1z}{1+z}\right)}
\ee

\be
\Omega_{\phi}(z)=\dfrac{1}{1+\alpha(1+z)^{-3(\omega_0+\omega_1)}\exp\left(\frac{3\omega_1z}{1+z}\right)}
\ee

\begin{figure*}[htb!]
\centering
\subfigure[]{\includegraphics[scale=0.39]{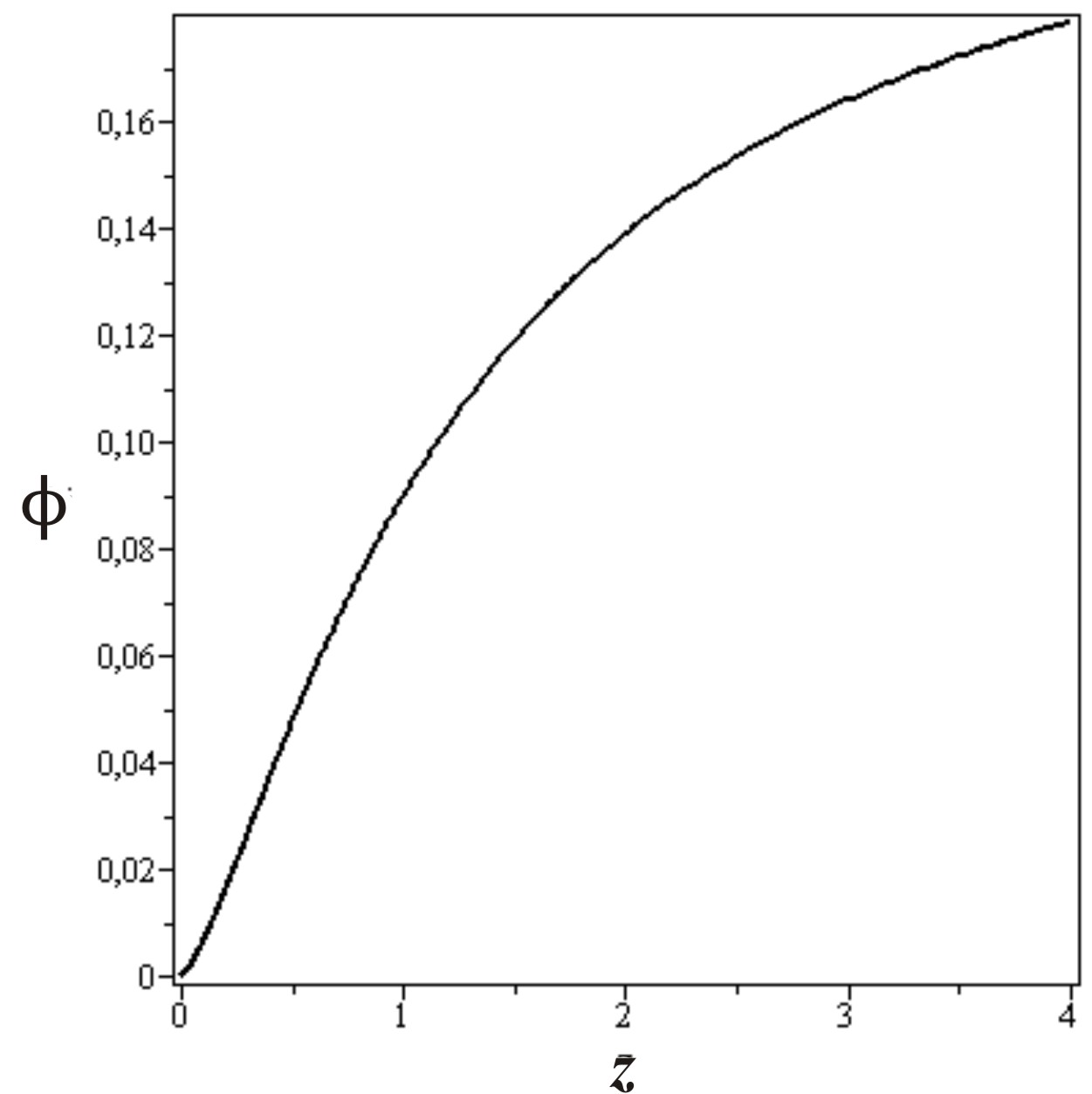}}\;\subfigure[]{\includegraphics[scale=0.39]{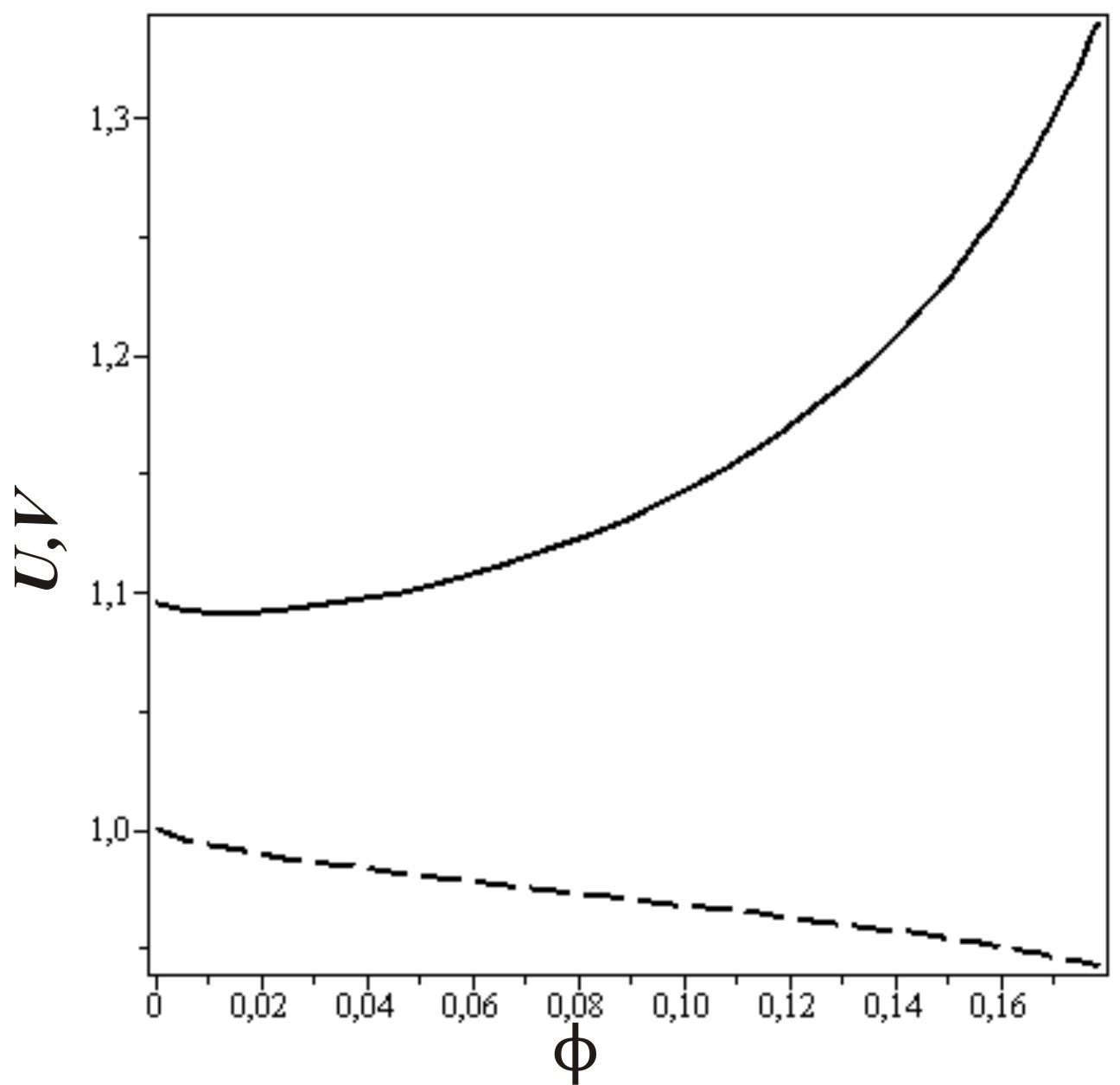}}\;\subfigure[]{\includegraphics[scale=0.39]{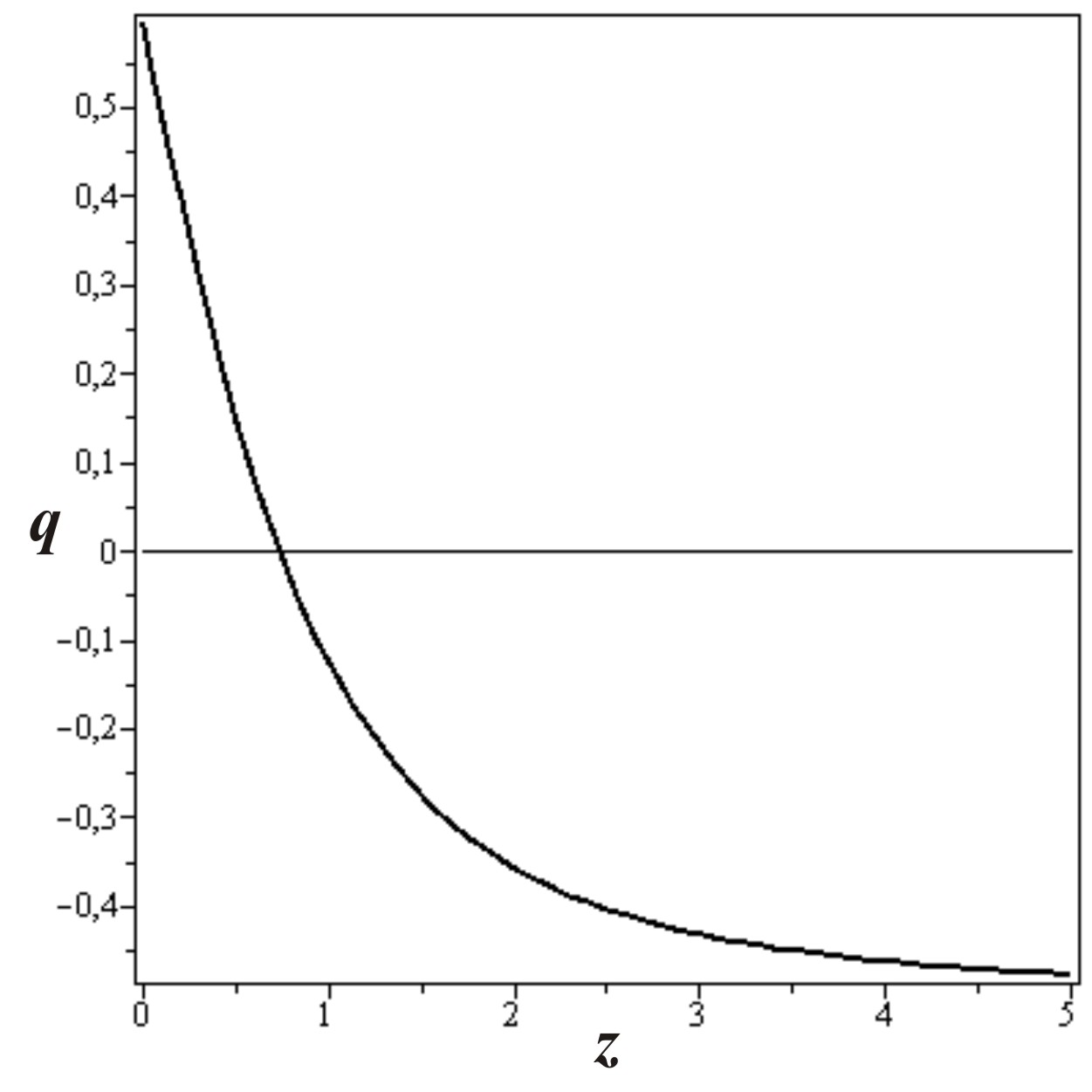}}\;\subfigure[]{\includegraphics[scale=0.39]{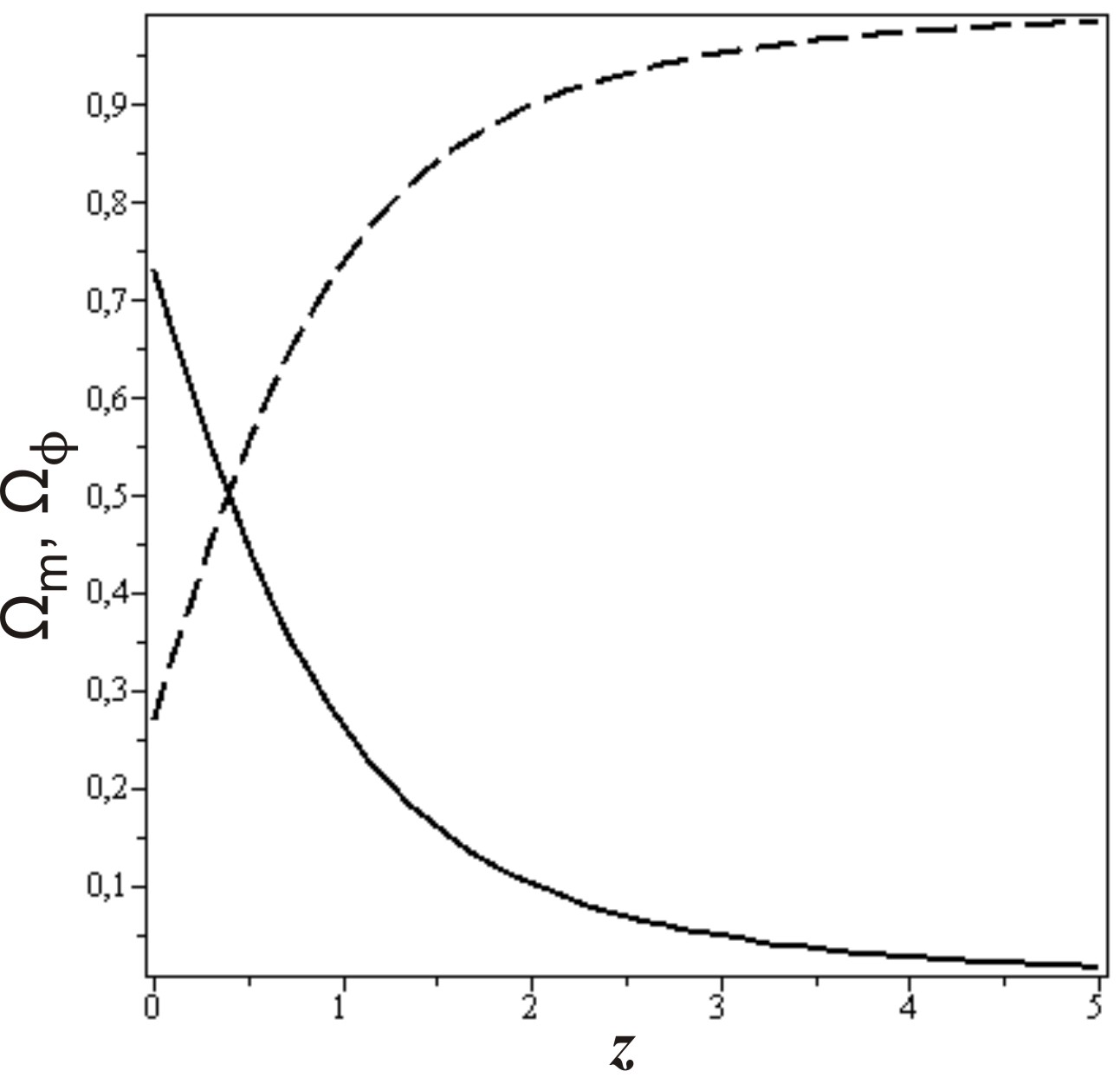}}
\caption{(a) Plot of $\phi$ as a function of $z$. (b) Plot of $V$ (solid curve) and $U$ (dashed curve) as a function of $\phi$. (c) Plot of $q$ as a function of $z$. (d) Plot of $\Omega_{\phi}$ (solid curve) and $\Omega_m$ (dashed curve) as a function of $z$. This is for CPL parametrization, with $\phi_0=0$, $H_0=1$, $\omega_0=-1$, $\omega_1=0.1$ and $\Omega_{m0}=0.27$.}
\label{fcpl}
\end{figure*}

Figure \ref{fcpl} shows the plots of (a) $\phi(z)$, (b) $V(\phi)$ and $U(\phi)$, (c) $q(z)$, (d) $\Omega_m$ and $\Omega_{\phi}$. Once again, the distinction between the models is evidenced in the graphs of $V$ and $U$, as well as the twin nature of these models requires that the curve of the acceleration parameter $q$ be the same for both cases.

\subsection{Barboza-Alcaniz (BA) parametrization}

The last example is the BA parametrization, proposed by Barboza and Alcaniz \cite{barboza,magana,yang}, which is represented by
\be
\omega_{\phi}(z)=\omega_0+\omega_1\dfrac{z(1+z)}{1+z^2}
\ee
In this case, the energy density of scalar field is
\be
\rho_{\phi}(z)=\dfrac{3}{2}H_0^2(1-\Omega_{m0})\left(1+z\right)^{3\left(1+\omega_0\right)}\left(1+z^2\right)^{\frac{3\omega_1}{2}}
\ee
The potentials of standard and modified cases are, respectively,
\ben
V(z)&=&\dfrac{3}{4}H_0^2(1-\Omega_{m0})\bigg(1-\omega_0-\nn
&-&\omega_1\frac{z(1+z)}{1+z^2}\bigg)\left(1+z\right)^{3(1+\omega_0)}\left(1+z^2\right)^{\frac{3\omega_1}{2}}
\label{v2}
\een

\ben
U(z)&=&\dfrac{1}{2}\bigg[4-6H_0^2(1-\Omega_{m0})\bigg(1+\omega_0+\nn
&+&\omega_1\frac{z(1+z)}{1+z^2}\bigg)\left(1+z\right)^{3(1+\omega_0)}\left(1+z^2\right)^{\frac{3\omega_1}{2}}\bigg]^{1/2}
\label{u2}
\een
The Hubble parameter evolves as follows
\ben
H^2(z)&=&H_0^2\bigg[\Omega_{m0}(1+z)^3+\nn
&+&(1-\Omega_{m0})(1+z)^{3(1+\omega_0)}\left(1+z^2\right)^{\frac{3\omega_1}{2}}\bigg]
\een
The acceleration parameter is then
\be
q(z)=-\dfrac{1}{2}-\dfrac{3}{2}\left[\dfrac{\omega_0+\omega_1\frac{z(1+z)}{1+z^2}}{1+\alpha(1+z)^{-3\omega_0}\left(1+z^2\right)^{-\frac{3\omega_1}{2}}}\right]
\ee
The density parameters of the matter and scalar field are, respectively,
\be
\Omega_m(z)=\dfrac{1}{1+\frac{1}{\alpha}(1+z)^{3\omega_0}\left(1+z^2\right)^{\frac{3\omega_1}{2}}}
\ee

\be
\Omega_{\phi}(z)=\dfrac{1}{1+\alpha(1+z)^{-3\omega_0}\left(1+z^2\right)^{-\frac{3\omega_1}{2}}}
\ee

\begin{figure*}[htb!]
\centering
\subfigure[]{\includegraphics[scale=0.39]{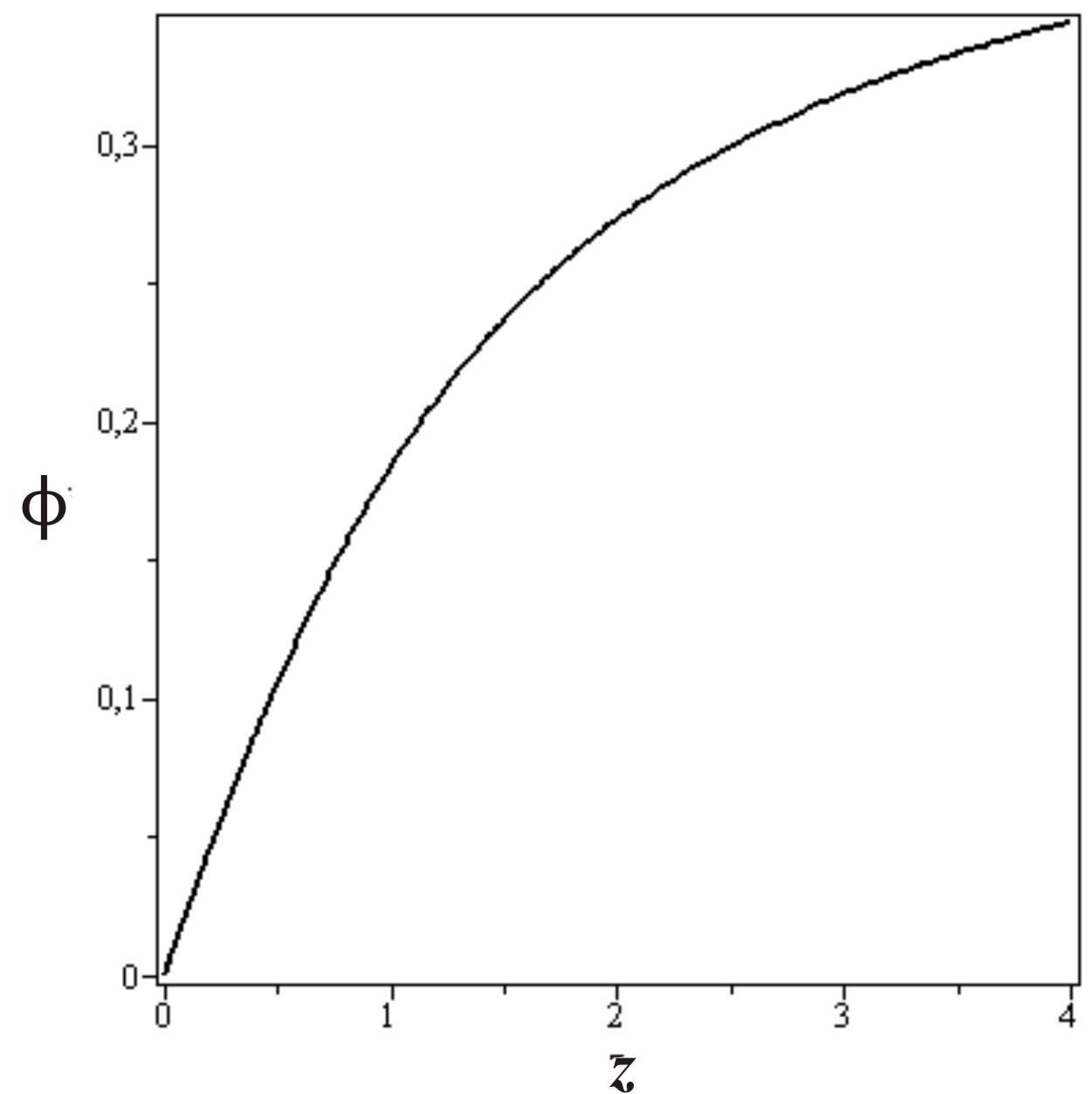}}\;\subfigure[]{\includegraphics[scale=0.39]{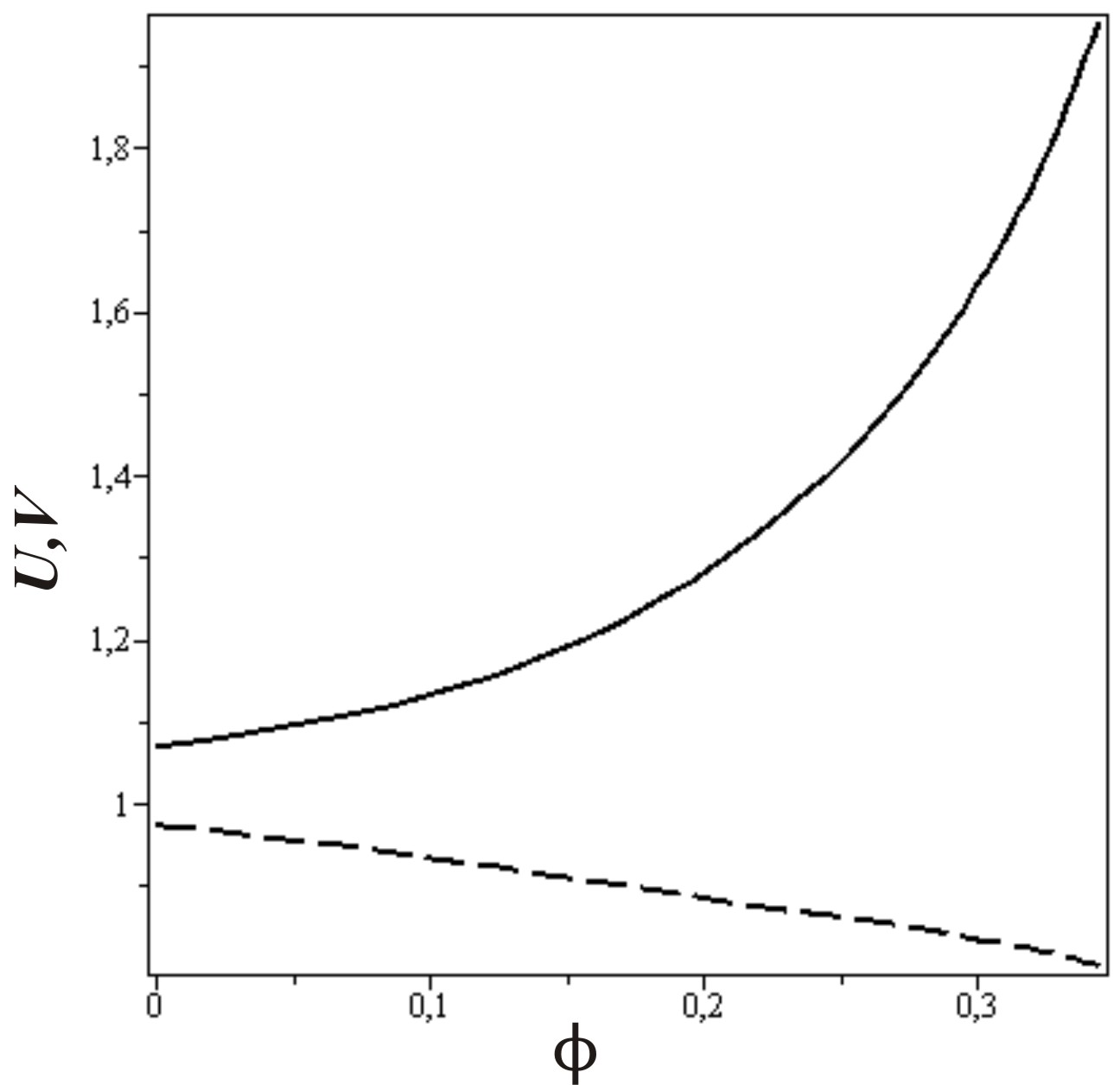}}\;\subfigure[]{\includegraphics[scale=0.39]{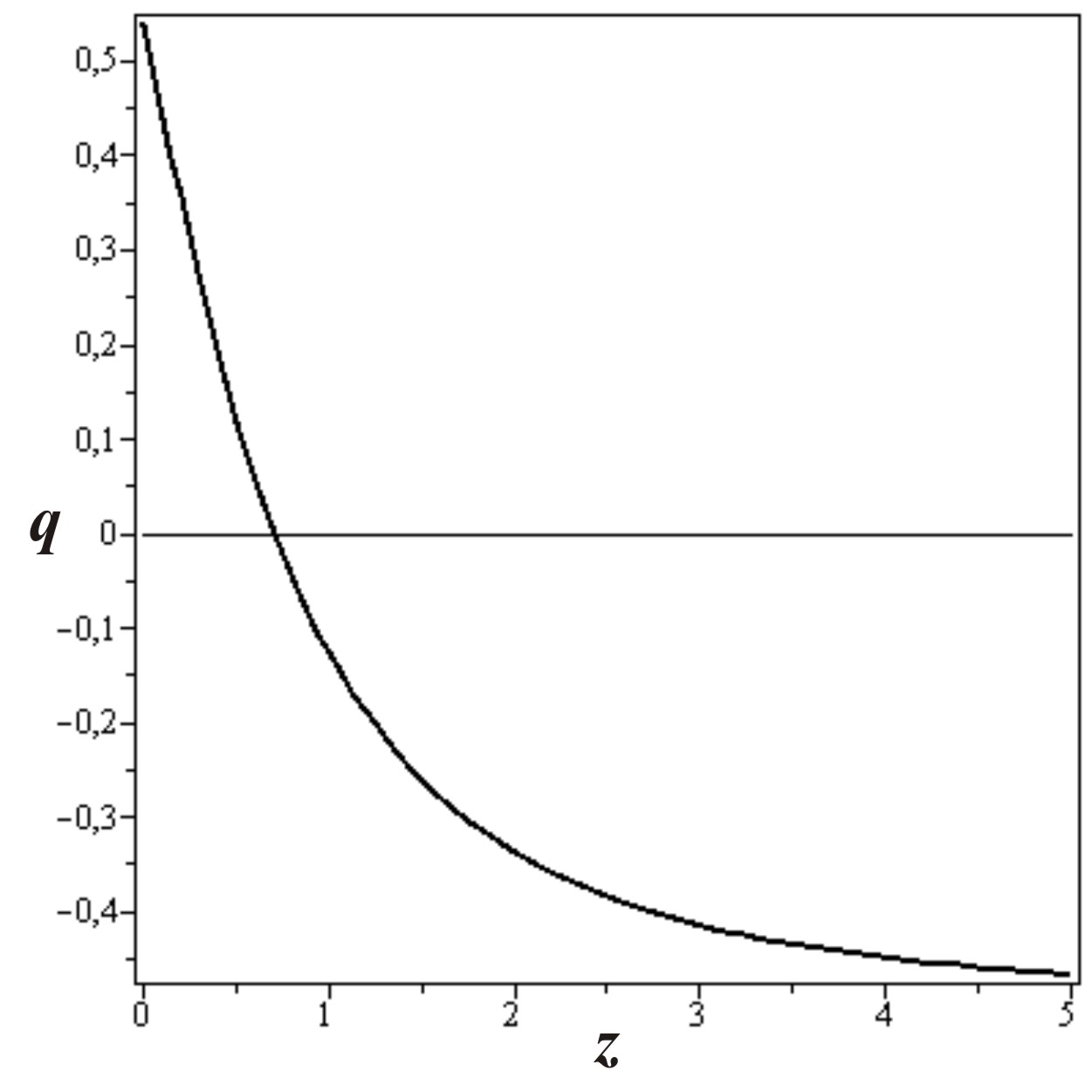}}\;\subfigure[]{\includegraphics[scale=0.39]{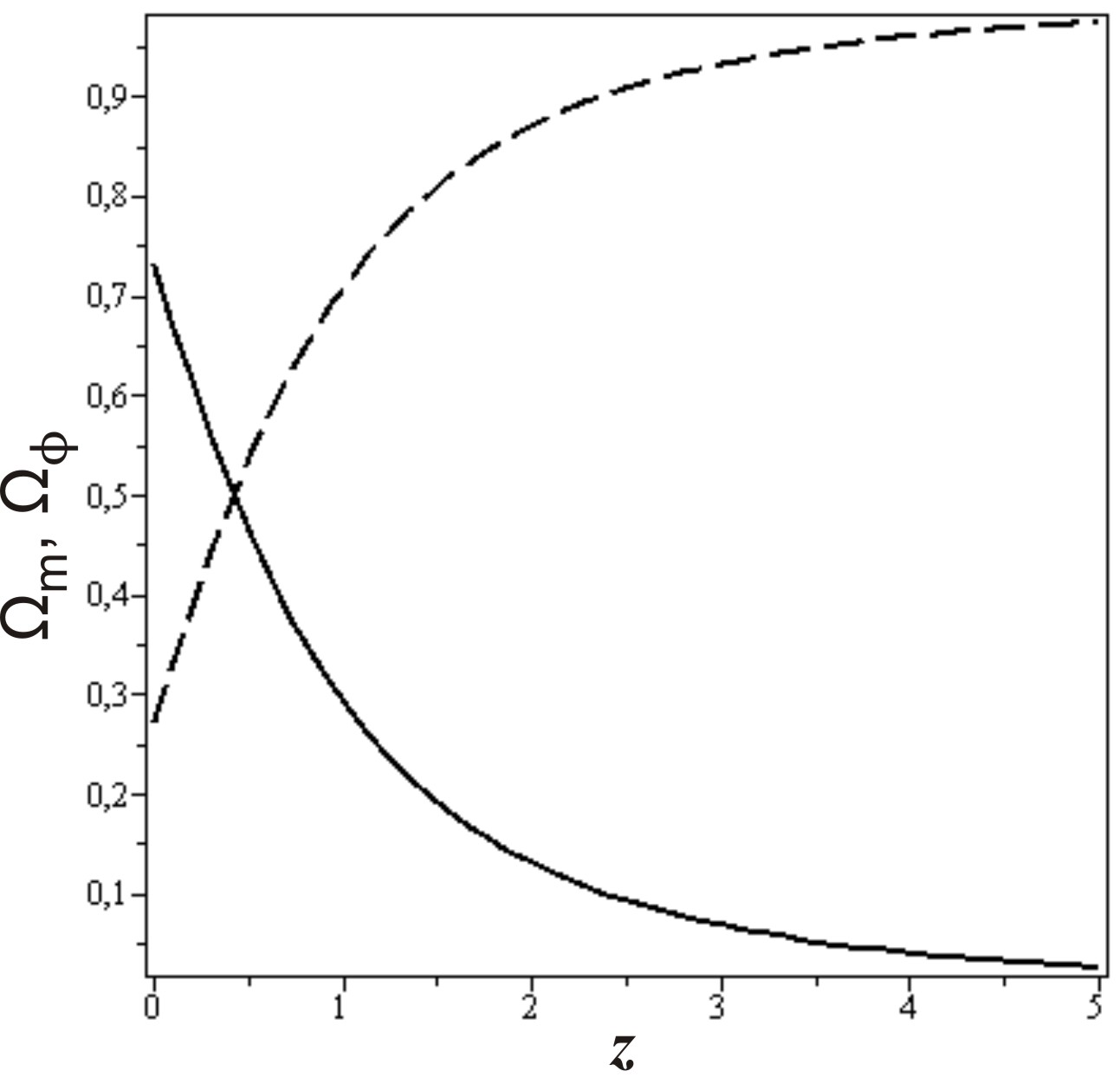}}
\caption{(a) Plot of $\phi$ as a function of $z$. (b) Plot of $V$ (solid curve) and $U$ (dashed curve) as a function of $\phi$. (c) Plot of $q$ as a function of $z$. (d) Plot of $\Omega_{\phi}$ (solid curve) and $\Omega_m$ (dashed curve) as a function of $z$. This is for BA parametrization, with $\phi_0=0$, $H_0=1$, $\omega_0=-0.95$, $\omega_1=0.1$ and $\Omega_{m0}=0.27$.}
\label{fba}
\end{figure*}

Figure \ref{fba} shows the plots of (a) $\phi(z)$, (b) $V(\phi)$ and $U(\phi)$, (c) $q(z)$, (d) $\Omega_m$ and $\Omega_{\phi}$. Discussion similar to previous ones can be performed around these graphs. The transition between $\Omega_m$ and $\Omega_{\phi}$ occurs again around $z\sim0.4$.

\section{Summary and conclusions}
\label{sandc}

In this work we studied the presence of twinlike models in FRW cosmology driven by a single real scalar field, in flat spacetime, from the study of some common parametrizations for the equation of state parameter $\omega(z)$. We showed that, regardless of the choice of $\omega(z)$, it is always possible to have models driven by standard and tachyonic dynamics with the same acceleration parameter, the same energy density and the same pressure.


\subsection*{Acknowledgments}
\label{ack}

We would like to thank CAPES for financial support and D. Bazeia for the fruitful discussions and suggestions.


\end{document}